# The Karl G. Jansky Very Large Array Sky Survey (VLASS). Data Products

Amy E. Kimball[a], Mark Lacy[b], Juergen Ott[a], John Tobin[b], Tierra Candelaria[a], Sergio Garza[a], Drew Medlin[a], and Steven T. Myers[a]

[a]National Radio Astronomy Observatory, 1011 Lopezville Rd, Socorro NM, USA
[b]National Radio Astronomy Observatory, 520 Edgemont Rd, Charlottesville VA, USA

## ABSTRACT

The VLA Sky Survey (VLASS) is a radio continuum survey of the entire sky visible to the VLA ($\delta > -40^o$) at ∼2.5 arcsecond resolution over 2–4 GHz with full polarization. Data processing and quality assurance have proven challenging due to limited computing and staff resources, and require innovative approaches. Although "Quick Look" images in Stokes I continuum were produced within a few weeks of observation, high accuracy images that utilize the full potential of the dataset are computationally intensive and take much longer to produce. Half of the sky will need correction for $w$-terms to account for sky curvature, increasing the computational cost per image by $\sim 100\times$. This problem will be tackled using GPUs and High-Throughput Computing (HTC) centers. Quality assurance of the tens of thousands of individual images has used a mix of algorithmic and machine learning techniques to automatically identify common artifacts and anomalies in the data. Archiving and dissemination of VLASS data products also pose a significant challenge.

## 1. INTRODUCTION

The Very Large Array (VLA) Sky Survey (VLASS) is a multi-epoch radio survey of the entire sky visible to the VLA, covering declinations $> -40^o$ (33,885 deg$^2$ of sky) over 2–4 GHz with full polarization. The survey and basic data product production are performed by the National Radio Astronomy Observatory (NRAO) as a service to the community, making the radio sky accessible to non-radio astronomers and citizen scientists with a resolution approaching that of modern optical and infrared sky surveys. Angular resolution varies across the sky but is $\sim 2.5$ arcseconds.

In this paper we briefly describe the VLASS observing strategy, discuss the basic VLASS data products produced at NRAO as well as quality assessment procedures, and give an overview of the current and expected data processing in order to complete the survey products. Reference is made to several memos in the NRAO library's VLASS Memo Series* when relevant.

## 2. OBSERVING

VLASS was originally envisioned as a three-epoch 2–4 GHz (S-band) survey to be performed in the VLA's B configuration (for most of the sky) and the hybrid BnA configuration (the southern sky). In the hybrid configuration, the east and west arms remain in B configuration while the northern arm is extended into the A configuration positions. The VLA was in the B/BnA configurations for approximately 5 months every 16 months, and half of the sky was observed in each configuration cycle with the complementary half being observed the next configuration cycle. During one configuration cycle, observations were performed in somewhat random order due to the VLA's dynamic scheduling process, and therefore the cadence between successive observations of the same region of the sky is $32 \pm 5$ months. It was eventually decided to observe half of the sky for a fourth time, in order to replace problematic products from Epoch 1.1 with Epoch 4.1 (see below), and thus VLASS

Further author information:
Amy E. Kimball, akimball@nrao.edu
*https://library.nrao.edu/vlass.shtml

ultimately observed the sky three and a half times over a period of eight and a half years from September 2017 through February 2026. The observing cycles are labeled Epoch 1.1 for the first observation of the first half of the sky, Epoch 1.2 for the first observation of the second half of the sky, Epoch 2.1 for the second observation of the first half of the sky, and so on. Table 1 shows the timeline for the VLASS observations. The survey used approximately 920 hours of VLA time in each observing cycle for a total of approximately 6500 hours.

Table 1. VLASS observing schedule for each half epoch.

| Half epoch | Observing dates |
|---|---|
| 1.1 | Sep 2017 – Feb 2018 |
| 1.2 | Mar 2019 – Jul 2019 |
| 2.1 | Jun 2020 – Nov 2020 |
| 2.2 | Sep 2021 – Feb 2022 |
| 3.1 | Jan 2023 – Jun 2023 |
| 3.2 | Apr 2024 – Oct 2024 |
| 4.1 | Aug 2025 – Feb 2026 |

While the nominal bandwidth for VLASS is 2–4 GHz, the actual bandwidth coverage with the correlator is 2.048 GHz in the range 1.965–4.013 GHz, with 16 spectral windows of 128 MHz each. This specific tuning was designed such that the bulk of the radio frequency interference (RFI) near 2.19 GHz (2178–2195 GHz) due to satellite downlinks was confined to a single spectral window (2.093–2.221 GHz).

Details of the survey design and observing strategy are provided in Ref.[1] In brief, the survey used an observing mode called "on-the-fly-mosaicking" (OTF), where the antennas are in a constant raster motion collecting data from chunks of the sky labeled as VLASS "tiles": the sky was divided into tiers from $-40^{\rm o}$ declination to $90^{\rm o}$ (each tier is 4 or 5 degrees tall in declination) with each tier divided into tiles of approximately $40\,{\rm deg}^2$ each; these tiles each represent an individual observing unit. VLASS tiles are labeled in the survey according to tier number and tile number. There are 32 tiers from declination $-40^{\rm o}$ to $+90^{\rm o}$. The second tile in the fourth tier is labeled T04t02 where the capital T stands for tier and the lowercase t stands for tile. The arrangement of tiles by RA/dec boundaries is described in VLASS Memo #7.

The method of OTF scanning is to scan in right ascension at constant declination. For most of the sky, the scanning rate is approximately 3.31 arcminutes per second. The integration time is 0.45s, and the phasecenter is updated every 2 integrations for a phasecenter separation of 2.98 arcmin. The separation between scanning rows in right ascension is 7.2 arcmin leading to a survey speed of 23.83 square degrees per hour in OTF mode. However, at lower declinations the scanning rate is slower (by a factor of almost 2 for declination of $-40^{\rm o}$) in order to reach a similar sensitivity of $\sim 120\mu$Jy in one epoch, accounting for an increase in system temperature from spillover from the over-illuminated secondary reflector when pointing at low elevations. A complex gain calibrator is observed in pointed mode approximately once every 15 minutes, leading to an overall survey speed of approximately 20 square degrees per hour including calibration overheads.

VLASS is the first full sky survey to be performed in OTF mode. As such, the OTF mode was still being commissioned at the time, and, as it turns out, errors were discovered in the antenna pointing data that affect Epoch 1.1 but were fixed before Epoch 1.2 observing began. As explained in detail in VLASS Memo #12, antennas with old-style Antenna Control Units (ACUs) (approximately 2/3 of the antennas during Epoch 1.1) were mis-pointed due to an error in the timing for commands sent to those ACUs. See Sec. 3 for discussion of the data products as affected by this pointing offset.

RFI posed a complication for VLASS due to strong S-band interference at the VLA. As discussed at the NRAO's RFI page[†] for the VLA, the bulk of S-band RFI in 2010 was due to satellite downlinks (2178–2195 GHz and 3700–4200 GHz) and Sirius/XM radio (2320–2350 GHz). The satellite downlinks were especially problematic, affecting the second lowest frequency spectral window (2.093–2.221 GHz) and the three highest frequency spectral windows (3.629–4.013 GHz), and most strongly interfering with observations near the Clarke Belt. To avoid the worst of the RFI, observations near $-5^{o}$ declination (the location of the Clarke Belt viewed from the VLA) were performed at especially high hour angles looking west. Additionally, over the 9 years of the survey an increasing number of images were affected by RFI due to, for example, Starlink and other satellites, and potentially increased use of smart devices by travelers along Highway 60 in New Mexico, which passes directly through the north arm of the B and BnA configurations of the VLA. For a discussion of how RFI is dealt with in VLASS images, see Sec. 4.

[†]http://go.nrao.edu/vla-rfi

## 3. NRAO DATA PRODUCTS

The collection of NRAO-provided "basic" VLASS data products consists of raw data and calibration tables (Sec. 3.1), "Quick Look" (QL) continuum images (Sec. 3.2), "Single Epoch" (SE) continuum images (SECIs) (Sec. 3.3), SE coarse cubes (Sec. 3.4), and cumulative images (Sec. 3.5). At the time of this writing, the raw data, calibration tables, and QL images are available for all observations. SECIs are being produced while coarse cubes and cumulative images are planned for the future. Coarse cube processing is set to begin in late 2026, while cumulative images will be produced once SECI processing has completed.

### 3.1 Raw and calibrated visibilities

As is the case for all VLA observations, raw VLASS data were immediately ingested into the NRAO data archive (see Sec. 6) and made available within minutes of observing completion. To calibrate the visibilities, the data were run through a VLASS-specific recipe of the standard VLA calibration pipeline[‡] integrated within the Common Astronomy Software Application (CASA).[2] Because the data were taken over many years, the versions of CASA and the pipeline varied over time but the procedure remained generally the same; different versions of CASA and the pipeline have been validated to ensure that they produce scientifically equivalent images. In short, the pipeline performs standard flux, bandpass, polarization and phase calibration and flags the target data for RFI. The consequential differences between the VLASS recipe and the standard recipe are: the VLASS recipe does not apply "quack" flagging (flagging the beginning of each scan), due to the use of OTF imaging, nor flag the edge channels of basebands or spectral windows; the VLASS recipe explicitly sets parameter `tbuff`—the time in seconds used to pad flagging commands in time—to be 0.225 (half of a VLASS integration) rather than the standard value of 1.5 integrations; the VLASS recipe explicitly limits the short solution interval in task `hifv_solint` to 0.45s (one VLASS integration). In addition, the VLASS calibration pipeline makes use of a task developed for VLASS known as `hifv_fixpointing`, which updates the pointing metadata for OTF data and applies polarization calibration using the `hifv_circfeedpolcal` task with polarization angle calibration information appropriate for the VLASS polarization angle calibrators (3C286 and 3C138). Details of the pipeline tasks can be found in the pipeline tasks reference manuals available at this link[§].

An additional calibration action developed for the VLASS pipeline (now available for more general use in the standard VLA pipeline) is the use of the switched power system at the VLA to correct for gain compression from strong signals (i.e., RFI). Gain compression refers to a deviation from linearity of the output-to-input response of the system. The parameter "$P_{\rm diff}$" is used to estimate the level of gain compression, where $P_{\rm diff}$ is the normalized power difference from when a calibrated noise signal is injected into the signal path at 10 Hz intervals; values $P_{\rm diff} < 1$ indicate some level of compression. The fix is essentially a division by the $P_{\rm diff}$ value as determined from spectral windows that are largely RFI free. NRAO staff determined that the division by $P_{\rm diff}$ is effective in calibrating for gain compression where $P_{\rm diff} \gtrsim 0.7$. Originally this fix was only applied to VLASS calibrations where compression was deemed to be worse than a 3% level ($0.7 < P_{\rm diff} < 0.97$); data with values $P_{\rm diff} < 0.7$ were flagged. It was later determined that applying the fix to non-compressed data did not harm the calibration, so in later epochs (and for all calibrations used for SE imaging), this fix has been applied to all calibrations using the pipeline task `hifv_syspower`.

For observations from Epoch 2.1 and earlier, one version of the calibration pipeline was used to produce QL images and a newer version is/was used to produce SE images. The earlier version of the pipeline, used for

[‡]https://science.nrao.edu/facilities/vla/data-processing/pipeline

[§]https://science.nrao.edu/facilities/vla/data-processing/pipeline/pipeline-version-history

QL images, extended the RFI flagging beyond what was necessary or desirable; this was fixed in newer versions of the pipeline. Affected observations were rerun through the newer calibration pipeline in preparation for SE imaging, and as a result there are many VLASS observations with more than one set of calibration products in the NRAO data archive (see Sec. 6).

### 3.2 Quick Look (QL) images

In order to provide images on a short timescale after observation, particularly to support time domain studies, the survey produced what are known as Quick Look (QL) images. QL images are Stokes $I$ continuum images averaged over the entire band, using a reference frequency of 3 GHz. These images are made with the mosaic gridder in CASA, have pixel sizes of 1 arcsec—so they do not sample the $\sim$2.5 arcsec beam particularly well—and were made without self-calibration or masking. As a result, these images could be produced relatively quickly, and were generally provided to the community within two weeks of observation: straightforward images could run through the VLASS CASA imaging pipeline within a day, whereas a small fraction of problematic images could take weeks to run through the pipeline (and potentially require reprocessing after flagging; see Sec. 4) and were provided to the community on a best efforts time scale.

Each QL image is slightly larger than one square degree ($\sim 3722 \times 3722$ pixels, with 1-arcsec pixels) but was created by first imaging approximately 4 square degrees of sky ($2^{\circ} \times 2^{\circ}$) and then selecting out approximately the central square degree (a "subimage") in order to avoid aliasing effects, and in order to clean sources outside the central degree that might otherwise cause sidelobes in the subimage. Each image was produced only using data from its own observation, and therefore images at the edge of a tile may cover a smaller region of sky than one square degree, and have higher noise at the edges because less data (from a single observation) is available at the tile edges. Images are identified by pre-defined phasecenters, with $\sim$40 phasecenters (images) per tile depending on the size of the tile. The subimage associated with phasecenter J061400−013000, for example, covers approximately one square degree centered on right ascension 6h14m00s and declination −01d30m00s. Image phasecenters are separated by one degree, and thus there is slight overlap at the subimage edges. In all, there are 35,500 image phasecenters across the 33,885 deg$^2$ of sky visible to VLASS.

Most QL images converged on the first imaging attempt, using the standard CASA `tclean` stopping criterion `nsigma`=4.5. `Nsigma` is a multiplicative factor for rms-based thresholding to determine when image cleaning has converged. Other common stopping criteria for VLASS QL images included `nmajor` (maximum number of major cycles), with a value of 220 used for VLASS, or `niter` (maximum number of minor cycles), with a value of used 20,000 for VLASS. The default `cycleniter` value was 500, limiting the number of minor cycles per major cycle to 500. For some parts of the sky that contain particularly bright sources, the default QL imaging pipeline did not clean sufficiently, but instead diverged. If an image diverged with the default parameters, a smaller `cycleniter` value was used and/or a smaller full image size. The smaller full image size could be used if the bright source was on the edge of the full image, to remove it from the data, or was well within the subimage, in order to clean faster when using a smaller `cycleniter`. The standard image size for QL full images was 7290×7290 pixels with 1 arcsec pixels. The smaller `imsize` value used for some images was 5760×5760 pixels. (In either case, the final subimage size is still one square degree.) Multiple combinations of `imsize` and `cycleniter` were attempted until `tclean` converged to create a suitable image. If `tclean` continued to diverge despite all attempts, or did not clean to the expected level, the QL image was rejected during the quality assurance process (see Sec. 4). Over the entire survey, $< 0.6\%$ of QL images were rejected. A description of CASA's `tclean` task and its input parameters can be found in the online CASA documentation.[¶]

QL image products include the images themselves and a corresponding rms sensitivity (noise) image. The published images have been corrected for the mosaic primary beam sensitivity. Details of the characterization of QL images can be found in VLASS Memo # 13. In short, tests were done to determine flux density accuracies by comparing the flux densities of phase calibrators that were observed in the survey with OTF observations and were also used for VLASS calibration from pointed observations. These tests show that, for most VLASS Epochs, the integrated flux densities are accurate within 3%, while peak intensities are low by about 8% due to residual phase errors. Due to the errors in antenna pointing positions during Epoch 1.1 (see Sec. 2), integrated flux densities for that half-epoch are low by about 10% while peak intensities are low by about 15%. At no time should flux densities above 1 Jy in Epoch 1.1 QL be used, due their to unreliability.

[¶]https://casadocs.readthedocs.io/en/stable/api/tt/casatasks.imaging.tclean.html

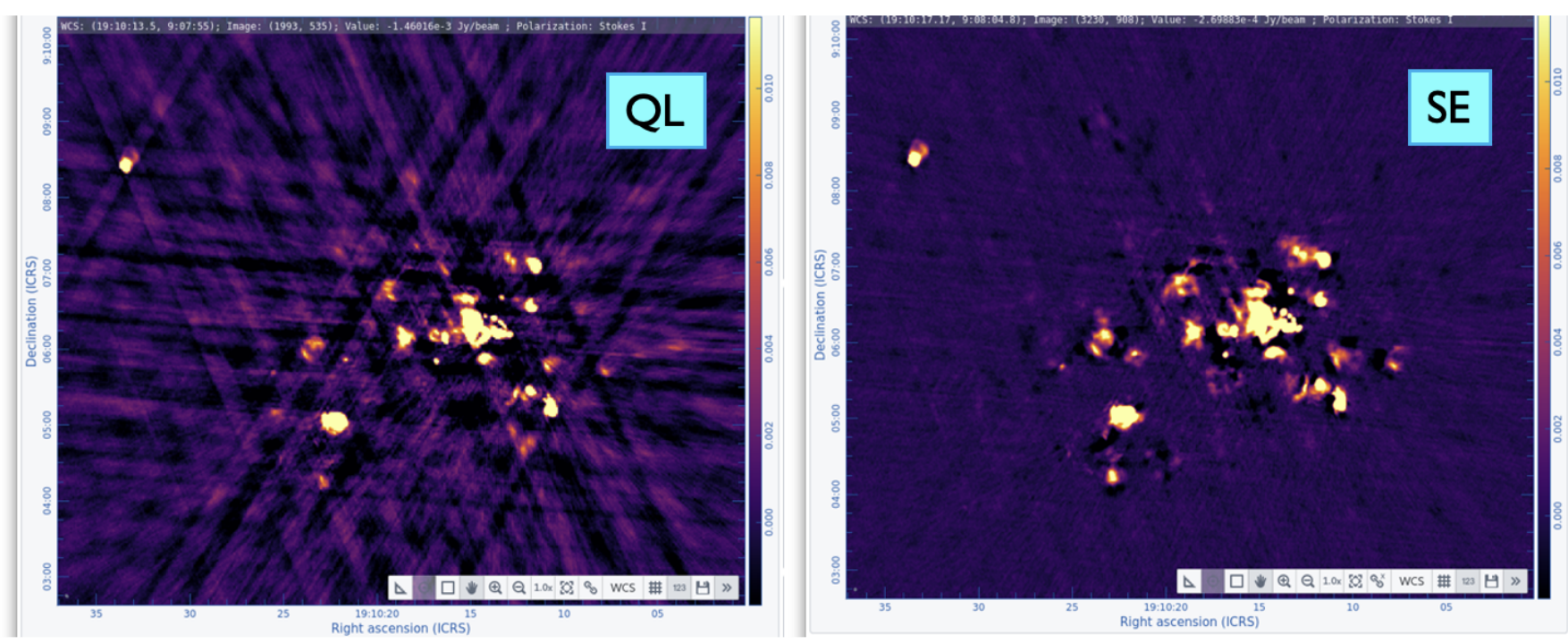

Figure 1. Comparison of a Quick Look (QL) image (left) and Single Epoch Continuum Image (SECI) for the same region of sky. The images shown here correspond to VLASS phasecenter T13t29.J191022+903000.

We note that QL image positions have been updated to correct for $w$-terms by adjusting the image center metadata by the appropriate offset. See Sec. 3.3 for a discussion of $w$-terms. Unfortunately, the QL images for Epoch 1 were initially ingested into the NRAO data archive before this error was identified and before the correction was made to the metadata. Therefore the Epoch 1.1 and Epoch 1.2 images, at the time of this writing, are slightly offset from the true sky in the official NRAO data archive. NRAO plans to correct the image metadata for images in the archive at some point in the future. However, cached versions of the images are also stored on disk at NRAO and are available for download at a different location; these cached versions have been corrected for $w$-terms, as have those available at the Canadian Astronomy Data Centre (CADC). Note that QL image positions for the rest of the epochs are correct in all locations. See Sec. 6 for details about how to access VLASS images.

### 3.3 Single Epoch continuum images (SECIs)

For each epoch of VLASS, NRAO is producing 35,500 of what are known as Single Epoch (SE) continuum images (SECIs) with identical phasecenters to those used for QL imaging. SECI subimages are approximately the same size as QL subimages; they are 6202 pixels on a side or 3721.2 arcsec with 0.6 arcsec pixels. The VLASS SECI pipeline recipe is more sophisticated than the QL recipe, with multiple rounds of imaging in order to incorporate self-calibration and masking. The pixel size in SECI is 0.6 arcsec, which better samples the interferometric resolution element ("clean beam") but results in much larger images in terms of number of pixels. The full image size used in `tclean` is 12500×12500 with a smaller `imsize` of 9600×9600 for some problematic images. SECIs use the same CASA `tclean` convergence criteria as QL images. Overall, the SECIs are much improved over QL due to the more sophisticated imaging, as shown for example in Figure 1.

SECI products include regular continuum images as well as spectral index images. Spectral index is defined as $\alpha$ where $S_\nu \propto \nu^\alpha$, $S_\nu$ is the flux density, and $\nu$ is frequency. Additionally, $\alpha = \mathrm{tt1}/\mathrm{tt0}$ where tt0 and tt1 are the zeroth and first order Taylor terms of a power-law index fit to the spectral shape at the VLASS reference frequency of 3 GHz. The products therefore include primary-beam-corrected tt0 and tt1 images as well as tt0 and tt1 flat noise images, and a precomputed spectral index "alpha" image along with an alpha error image.

SE products also include image-specific catalogs made with PyBDSF.[3] The default PyBDSF values are used for the parameters `thresh_isl` and `thresh_pix`. The values for `rms_box` are (60,20) meaning that the rms is calculated adaptively every 20 pixels (12 arcsec) using a box 60 pixels (36 arcsec) on a side. Around bright sources, the values used are `rms_box_bright` = (20,6) corresponding to a box size of 20 pixels (12 arcsec) calculated every 6 pixels (3.6 arcsec). Because subimages are slightly larger than 1 square degree, and due to the of the curvature of the sky, there is some overlap at the image edges and thus also overlap in the catalogs made from individual images. See Sec. 3.6 for a discussion of combined catalogs that are available.

SE continuum imaging is a considerably greater challenge than QL imaging. In continuum, the combination of wide field and wide-band imaging mean that the curvature of the sky ($w$-terms) cannot be ignored, or easily corrected, for observations taken at low elevation. This was first determined for VLASS when an offset of $\approx 0.5 - 1.5$ arcsec was seen between source positions in the QL images and reference positions from the ESA Gaia astrometric space mission[4] and Very Long Baseline Interferometry (VLBI).[5] These offsets were eventually established to be a manifestation of the neglect of $w$-terms in the imaging of the OTF data. Typically, in a single pointing image, the neglect of $w$-terms manifests as a distortion of source structure in the image (e.g., ref.[6]) whose magnitude depends on observing frequency, the tangent of the zenith distance of the observations, and distance of the source(s) from the image phasecenter. In the OTF case, with many hundreds of pointings making up an image, the first order contributions to these distortions are averaged out, but there remains a frequency dependent position offset that is proportional to the square of the size of the primary beam and the tangent of the zenith distance.

Consider an observation of the sky made in a direction corresponding to the unit vector $\mathbf{s} = (\alpha, \delta)$, where $\alpha$ is the Right Ascension and $\delta$ the Declination. Using the standard conventions, let $\mathbf{u}$, $\mathbf{v}$, $\mathbf{w}$ be the unit vectors in the coordinate frame defined by the telescope baselines, with $\mathbf{w}$ perpendicular to the $uv$-plane, which is itself parallel to the tangent plane on the sky. Let $l$, $m$, and $n$ be the direction cosines to $\mathbf{s}$ in the $u$, $v$, $w$ coordinate system, i.e., $l = \mathbf{s} \cdot \mathbf{u}$, $m = \mathbf{s} \cdot \mathbf{v}$, $n = \mathbf{s} \cdot \mathbf{w}$, and $l^2 + m^2 + n^2 = 1$. Then it can be shown that the visibilities, $V$, can be written as the 3D Fourier transform of the sky brightness $I$ and primary beam correction $P$:[7]

$$V(u, v, w) = \int\int I(l, m) P(l, m) \mathrm{e}^{-2\pi i (ul + vm + nl)} \frac{\mathrm{d}l\, \mathrm{d}m}{\sqrt{1 - l^2 - m^2}}. \tag{1}$$

If we define $F$ as the 3D Fourier transform of the visibilities, then

$$F(l, m, n) = \int\int\int V(u, v, w) \mathrm{e}^{2\pi i (ul + vm + wn)} \mathrm{d}u\, \mathrm{d}v\, \mathrm{d}w. \tag{2}$$

The phase (equivalent to a position error if not corrected for) introduced by the $w$-term is thus $2\pi n w = 2\pi\sqrt{1 - l^2 - m^2} w \sim \pi(l^2 + m^2) w$.

The differing projection of the $u, v$ plane across the primary beam is an important aspect of this analysis in the case of wide-field imaging.[7] The scaling of the $u, v$ axes is determined by the local tangent plane. For observations off zenith, the projected lengths of the baselines are foreshortened by an amount depending on the hour angle and declination of the phase center. The mosaicking algorithm (in the absence of $w$-term corrections) convolves each point in the $uv$ plane by the Fourier transform of the primary beam. Thus any point in the image plane can be thought of as being comprised of containing contributions from the sky image multiplied by the primary beam response. Points in the primary-beam multiplied sky image farther from zenith will thus have a slightly differently-scaled $uv$-plane than those closer to zenith, resulting in a nonlinear scale variation of the coordinates across the primary beam, and a shift in the source position when the image is reconstructed.

Mathematically, the analysis is as follows. It can be shown[8] that at any given instant, $w$ can be eliminated from Equation (1) by rewriting it in terms of projected $l$ and $m$ values $l'$ and $m'$. (Snapshot imaging is therefore immune to the effects of $w$ provided the requisite distortion is removed from the image. Thus, these effects do not show up in surveys made up of distortion-corrected images from pointed observations combined in the image plane, for example the Faint Images of the Radio Sky at Twenty centimeters (FIRST) survey:[9]

$$V(u, v, w) = \int\int I(l, m) P(l, m) \mathrm{e}^{-2\pi i (ul' + vm')} \frac{\mathrm{d}l\, \mathrm{d}m}{\sqrt{1 - l^2 - m^2}}. \tag{3}$$

Cornwell, Golap & Bhatnagar (2003, EVLA Memo 67; 2008) give the relationship between $l$ and $m$ coordinates and the 2D projection, $l'$ and $m'$, in terms of the zenith distance, $\zeta$, and the parallactic angle, $\chi$: [‖]

$$l' = l + \tan\zeta \sin\chi \left( \sqrt{1 - l^2 - m^2} - 1 \right) \tag{4}$$

[‖]The parallactic angle is given by:

$$\chi = \arctan(\sin H / (\cos\delta \tan\Lambda - \sin\delta \cos H)$$

$$m' = m + \tan\zeta\,\cos\chi\,\left(\sqrt{1-l^2-m^2}-1\right). \tag{5}$$

Consider a source towards the edge of the primary beam, an angle $\epsilon$ from the phase center such that $\epsilon^2 = l^2 + m^2$ and $\epsilon^2 << 1$. Then, expanding up to 2nd order in $\epsilon$,

$$l' - l \approx -\frac{\epsilon^2}{2}\tan\zeta\,\sin\chi \tag{6}$$

$$m' - m \approx -\frac{\epsilon^2}{2}\tan\zeta\,\cos\chi. \tag{7}$$

In practice, we can treat $\epsilon$ as a free parameter of order the primary beam width, as it is not clear exactly how the mosaicking will weight data from different parts of the primary beam, particularly for the wide bandwidth used by VLASS. Comparing to the simulations and analysis of VLASS tiles at extreme hour angle and declination, we find that $\epsilon \approx 5.4$ arcmin.

In the QL imaging the offset above was corrected for by adjusting the image center metadata by the calculated value of the offset (see VLASS Memo #14). For the SECI imaging, tiles closer to zenith (i.e., with a smaller zenith distance) are produced using the mosaic gridder in CASA to grid the uv-data onto a 2D grid and apply the same position correction as for the QL images. At high zenith distances, however, the effect on the synthesized beamshape and, especially, on the in-band spectral index (due to the frequency dependence of the $w$-term distortion) mean that imaging using $w$-terms is required to meet the survey requirements. The criteria used to determine whether or not the images require $w$-term imaging are based on the zenith distances to the phasecenters for images in each tile: for tiles that were observed with a median zenith distance $< 45^o$ and a maximum zenith distance of $< 50^o$, the mosaic gridder is being used. Tiles that fail to satisfy either one of these criteria are being set aside for processing with $w$-terms.

There was a large delay in the start of SECI processing due to time spent figuring out the appropriate algorithms required to produce SECIs in CASA, and as a result the images are being produced out of chronological order. At the time of this writing, mosaic-gridder SECIs have been produced fully for Epochs 2 and 3, and are now being produced for Epoch 1.2, after which the mosaic-gridder SECIs will be produced for Epoch 4.1. However, there is no plan to produce SECIs for Epoch 1.1: the computing required to correct for the pointing offsets (see Sec. 2) is prohibitive. Epoch 1.1 SECI products are being replaced by Epoch 4.1 products.

The SECI images produced by the mosaic gridder have been analyzed for flux density, spectral index and positional accuracy. Figure 2 shows the flux density of calibrators measured during the observations (using pointed scans) with the flux densities recovered from the mosaics using pyBDSF (for which we used the Isl_Total_fluxes; see Section 3.6). For calibrators with multiple observations we picked the observation closest in time to the epoch of the mosaic image (within 5 days). For the 77 calibration observations within one day of the corresponding calibrator image in the Epoch 2.1 SECIs made with the mosaic gridder, we find a mean difference of 0.53 ±0.17% and a scatter (scaled median absolute deviation from the median (MAD)) of 1.7%, indicating no strong systematic differences in flux density between the mosaicked VLASS images and pointed observations. For the spectral indices, we do see a mean systematic offset of 0.09 ±0.01 between the mosaic and flux calibrator images, together with a scaled MAD of 0.08 (Fig. 3). We believe that the offset may be due to the way the primary beam correction is handled, using the mean frequency primary beam rather than one that varies with frequency. Even with the offset, the combined offset plus scatter remains within the survey requirement of ±0.2 on the spectral index accuracy. For positions (after applying the $w$-term correction) the RMS position difference between VLASS and the VLBI sources in the Radio Fundamental Catalog[5] is 0.2 arcsec in right ascension and declination (Fig. 4).

---

where $H$ is the hour angle and $\Lambda$ is the latitude of the telescope, and the zenith distance is:

$$\zeta = \arccos\left(\sin\Lambda\sin\delta + \cos\Lambda\cos\delta\cos H\right).$$

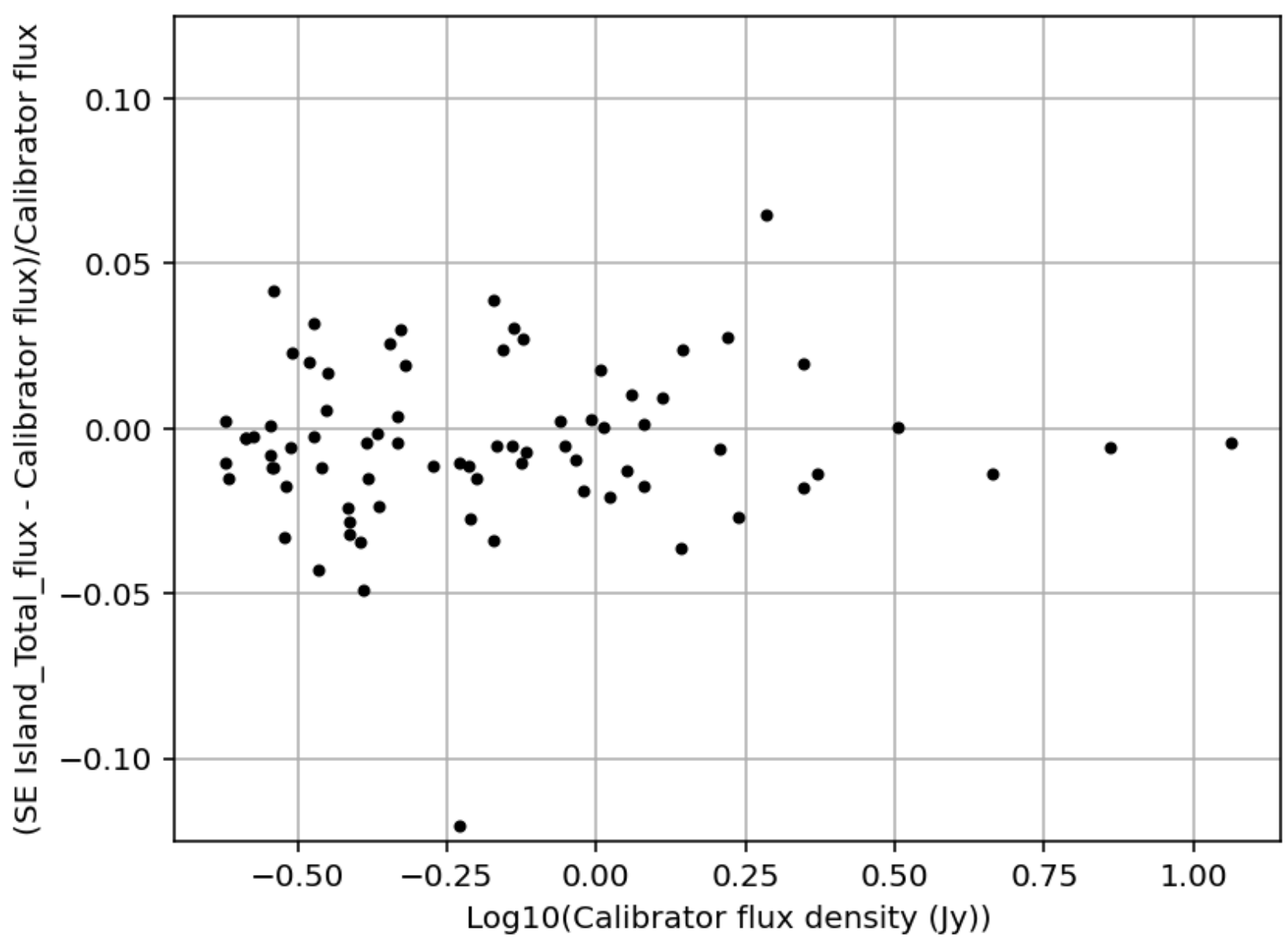

Figure 2. Comparison of the measured Island_Total_flux in the VLASS2.1 SECI mosaic gridder images to the calibrator fluxes measured using the standard gridder in the pipeline.

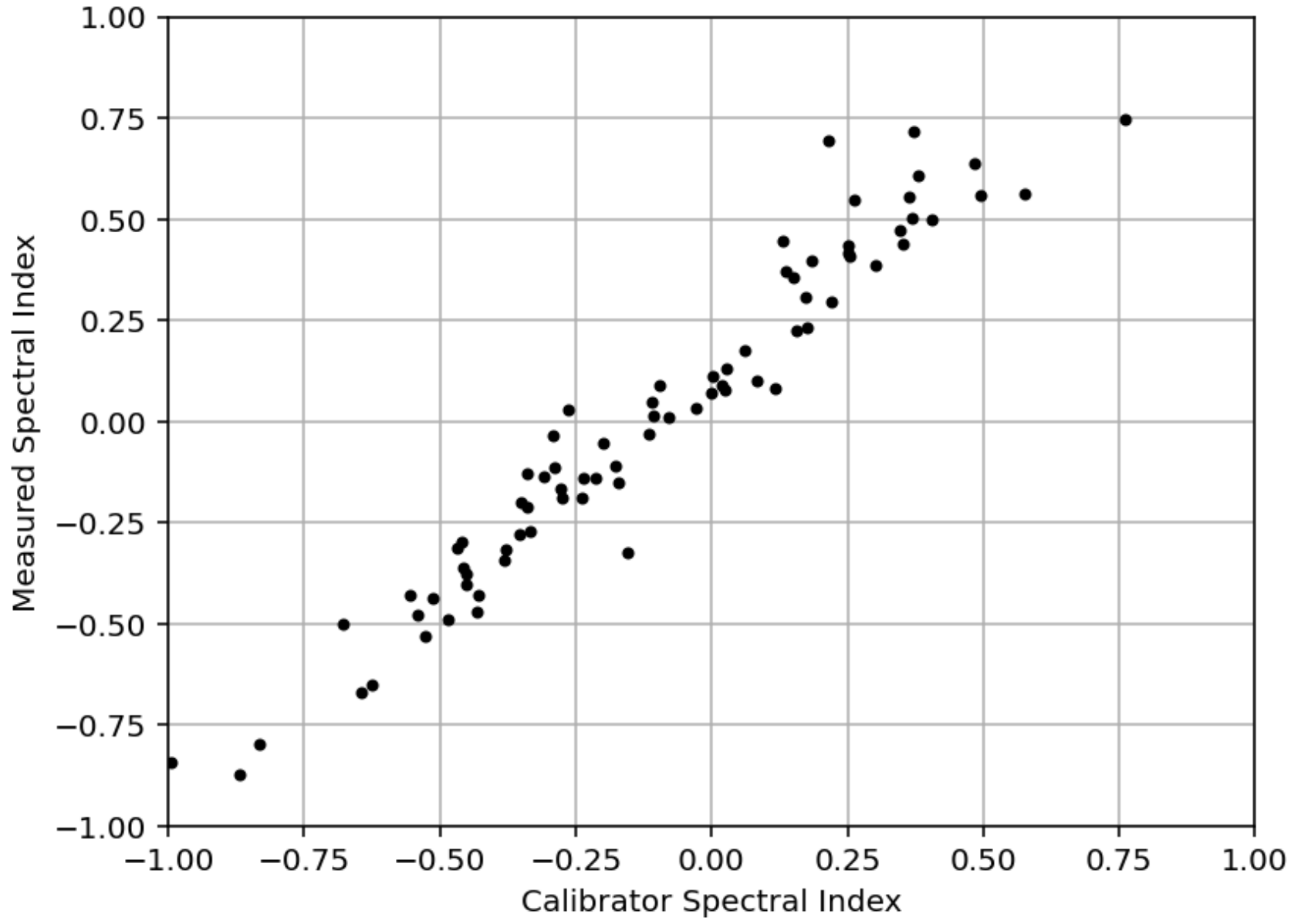

Figure 3. Comparison of the measured spectral index in the Epoch 2.1 SECI mosaic gridder images (vertical axis) to the calibrator spectral indices measured using the standard gridder in the pipeline (horizontal axis).

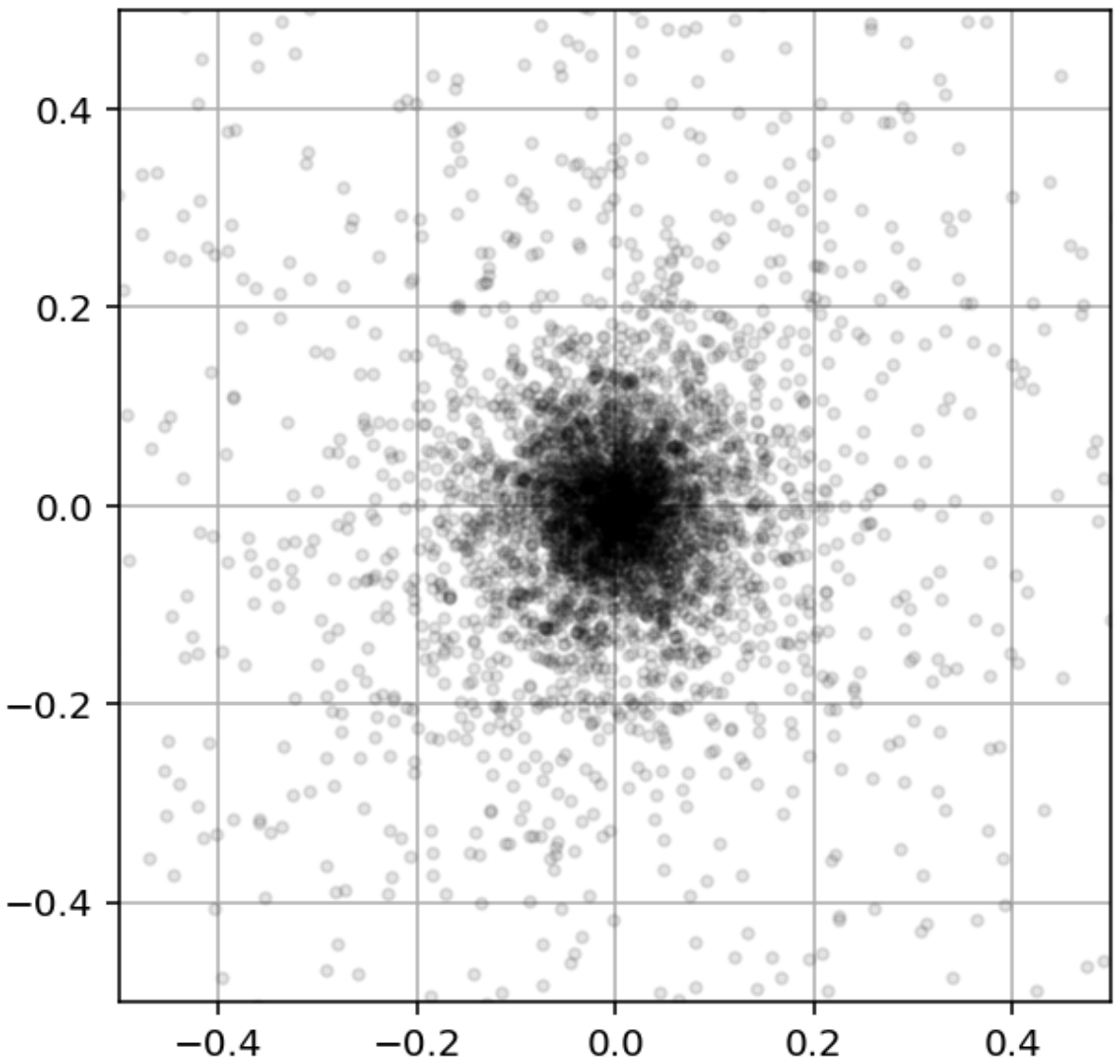

Figure 4. Comparison of the measured position in the Epoch 2.1 SECI mosaic gridder images to VLBI positions.

### 3.4 Single Epoch coarse cubes

Single Epoch "coarse cubes" will be created using a specialized version of the VLASS imaging pipeline, for use in Faraday rotation measure synthesis (e.g., ref.[10]). Each plane of the cube will represent the continuum averaged over 128 GHz of bandwidth—corresponding to a single spectral window—yielding 16 planes in frequency. These will also be cubes in polarization, with planes created for the I, Q, and U Stokes parameters. Thus each image phasecenter will correspond to a cube with up to 16 planes in frequency and each frequency having 3 Stokes planes. NRAO is also producing band-averaged V-plane images. The mosaic gridder will be used, with position corrections per frequency channel to account for sky curvature as discussed in Sec. 3.3. Coarse cube images will not include spectral index images; spectral indices can be measured by calculating flux density as a function of frequency across the Stokes I planes. The pixel size of these images will be 1 arcsec, rather than the 0.6 arcsec pixel size of the SECIs. As discussed in VLASS Memo # 24, the larger pixel sizes reduces the memory footprint of each job, allowing a higher computing throughput, and has minimal science impact due to the fact that current polarization analysis algorithms require the input images to be convolved to a common beam (i.e., the lowest frequency's beam, which is the largest). A 2-arcsecond taper will be applied in the $uv$-plane to avoid undersampling the highest frequency channels. Final coarse cube products will include the images themselves and a corresponding noise image for each plane.

Each coarse cube will have up to 16 planes in frequency (with 3 Stokes planes per frequency), but some number of frequency planes will be rejected due to poor quality. There are two main criteria for rejection: flagging level and beam size. If too much of the data are flagged in a particular plane, the pipeline will not image that plane. While the specific thresholds may change as NRAO learns more about coarse cube quality, the starting criterion is that a plane will be rejected in the pre-imaging round (and not imaged) if $> 12$ fields in the image are flagged above the 90% level. All other planes will be imaged by the pipeline. However, two additional flagging critera will be applied after imaging. First, a plane will be rejected by the pipeline if the flagging across all visibilities for that spectral window is above a certain threshold, with an expected threshold value of 80%. Second, if after imaging the beam major or minor axis for any plane is more than 20% (expected threshold) different from the median beam (across all spectral windows) scaled to the appropriate frequency, the plane will be rejected; a distorted beam is typically due to flagging in that specific frequency plane. Rejected planes that have been imaged will not be used in the pipeline's spectral index analysis; they may still be archived, in which case they will be flagged in the image header as rejected. After the pipeline run, a final decision for accepting or rejecting planes will be made during the quality assessment (QA) procedure.

### 3.5 Combined Epoch images

When three epochs of the survey have been processed as SECIs and SE coarse cubes, NRAO will produce cumulative or "combined epoch" images via simple image-plane combination. Tests reported in VLASS Memo #21 show that the results from image plane combination are similar to those resulting from joint imaging using the combined visibilities. There is a strong preference for image plane combination over joint deconvolution due to computational cost: joint deconvolution would require the use of $w$-term corrections over the entire survey due to the different elevations of tile observations across different epochs. Additionally, variable sources are problematic for joint deconvolution because the differences in flux density mean a highly variable source cannot be accurately cleaned with currently available algorithms.

The combined epoch images will be produced via convolution of the input images to a common beam that encompasses the input beams, as discussed in VLASS Memo #26; the combined image will be an RMS-weighted mean of the convolved input images.

### 3.6 Catalogs

The SECI catalogs represent the reference catalog products from the survey. The individual image catalogs described in Section 3.3 were combined and duplicate entries removed by matching components within 1 arcsecond and rejecting the component with the lower signal-to-noise (defined as the peak flux density divided by the pyBDSF island RMS). The flux density units in the catalogs are mJy and source size parameters are in arcseconds. Spectral indices for each component were measured by calculating the average of a 3x3 pixel grid of the spectral index map around the peak pixel of the component.

We note that very bright ($\gtrsim$ 1Jy) VLASS sources are often overfit by pyBDSF and the Island Total Flux (Isl_Total_flux) is generally more reliable than either the peak or total component flux densities. At lower signal-to-noise ratios, the tendency to overfit is much reduced and the total and peak fitted flux densities are preferred, as the island flux is affected by the threshold used by pyBDSF.

The component catalogs are then analyzed to find DRAGNS - Double Radio Active Galactic Nuclei.[11] Resolved (major axis >3 arcsec) components with flux densities $>$ 1.2mJy (lower than the 3 mJy limit used in Ref.[11] due to the higher quality of the SE images) are paired with their nearest neighbors that also satisfy those criteria. Candidate doubles are then selected using a metric that combines the angular separation and the misalignment of the position angles of the components (equations 3 and 4 of Ref.[11]). Core components are then identified according the criteria in section 2.3 of Ref.[11] The list of DRAGNs is then merged with the original list of single, compact sources and the components of the DRAGNs removed. Counterparts in the Wide-field Infrared Survey Explorer (*WISE*) catalog are then identified for both the doubles and compact/core components through likelihood ratio matching (again, following Ref.[11]). As a final step, a match is performed to the NASA Extragalactic Database (NED) using a cone search within 1.5-arcsec and, if a match is found, the NED preferred name for the source and, if available, redshift and redshift type flag (e.g. spectroscopic or photometric) written into the catalog.

In addition to the SECI catalogs, component catalogs for QL images in Epochs 1 and 2 have been made by the CIRADA collaboration.[12] similar catalogs for Epochs 3 and 4 and for a median stack of the images for Epochs 1, 2, and 3 have been made by NRAO.[**]

## 4. QUALITY ASSESSMENT

Quality assessment (QA) practices have varied over the course of the survey as NRAO has learned more about the data products, and has developed more automated procedures. This section explains the settled version of quality assessment for calibration and continuum imaging. Procedures are expected to be similar for coarse cubes and cumulative images.

[**] catalogs are available from https://vlass-dl.nrao.edu/vlass/quicklook/catalogs and https://vlass-dl.nrao.edu/vlass/se_continuum_imaging/catalogs

The initial goal for QA of VLASS calibrations was to determine whether the observation was successful or if any tile needed to be re-observed due to, e.g., bad weather or correlator issues. If re-observation was needed, the original observation of the affected tile was discarded, and a later successful observation was used for calibration and imaging instead. For successful observations, the calibration QA process served to determine if the default pipeline run was acceptable; often (50–60% of images) a single pipeline run with default parameters was sufficient. However, there were times when the QA process would identify poor data that were mistakenly highly weighted or would determine that the pipeline made a poor choice of reference antenna. In the former case, the data were flagged and the pipeline rerun. In the latter case, the pipeline was re-run with the parameter "refantignore" in the relevant pipeline tasks in order to choose a different reference antenna. Occasionally, phase jumps (strong variation in phase solutions with time) were identified on a misbehaving antenna, in which case those scans on either side of the phase jump would be flagged for the affected antenna. Another reason the pipeline would be re-run in the early half-epochs was to correct for gain compression (see Sec. 3.1); the fix for gain compression was later applied by default for all calibration runs (including all calibrations used for SE imaging). Finally, as noted in Sec. 3.1, the original version of the pipeline used for early half-epochs was extending the default flagging (i.e., extending the RFI flagging in time and frequency) more than necessary. Thus for observations from Epoch 2.1 and earlier, a second pipeline run with a newer pipeline version is being used for SE calibration. For later half-epochs, the same calibration (with the newer pipeline, with gain compression fix applied) is being used for both QL and SE imaging.

The QA process for imaging has developed over the lifetime of the survey. Initially, for Epochs 1 and 2, every QL image was examined visually by NRAO staff for quality and to identify possible artifacts which required flagging data before re-imaging. This visual examination was tedious and time-consuming, so NRAO staff developed a more automated procedure to examine image statistics and identify artifacts. The automated procedure (nicknamed "Otto") uses image statistics (see next paragraph) to determine if an image was cleaned to a sufficient level, and whether the image contained possible artifacts (see Sec. 4.1). Images cleaned to a sufficient level with no artifacts were automatically accepted for archiving and release to the public. Images which initially were not sufficiently cleaned were automatically reprocessed in `tclean` with updated imaging parameters as discussed in Section 3.2; the newer version of the image could be automatically accepted if subsequent cleaning was sufficient. Images which did not sufficiently clean after multiple attempts, or which contain certain artifacts, are examined visually. Approximately 80% of images are processed—reprocessed if necessary—and accepted automatically by Otto, while the remaining images require manual investigation for potential flagging due to artifacts, or poor deconvolution after multiple imaging attempts.

The image statistic used to determine image quality is the peak in the residual image divided by the median absolute deviation (MAD) multiplied by 1.4826 such that the scaled-MAD is equivalent to the root-mean-square (RMS) for a Gaussian distribution. The ratio of peak to scaled-MAD was measured both in the full image (2 deg $\times$ 2 deg) and the $\sim$1 square degree subimage (see Sec. 3.2). Recall that the subimage is the final published image. A QL image was deemed to be sufficiently cleaned for QA acceptance if the ratio of peak to scaled-MAD within the full image was $< 20$ and the ratio within the subimage was $< 6$. If these criteria were not met, a new version of the image was made with different input parameter values for the `cycleniter` and/or `imsize` parameters in `tclean` (see Sec. 3.2). If multiple imaging attempts were made and the peak to scaled-MAD reached $< 10$ in the subimage, the image could be deemed acceptable. All images that did not meet these criteria were examined manually. QL images were rejected if they did not meet the statistical criteria for acceptance. QL images were also rejected if the restoring beam was abnormally small (due to errors in `tclean`) or abnormally large (e.g., major axis $> 6$ arcsec).

The QA process is the same for SE continuum imaging, but has an additional component that is more subjective. Some additional SECIs are deemed acceptable if they do not meet the statistical criteria or the beam size criteria as long as the imaging converged in `tclean` and there are no egregious artifacts in the image (see Sec. 4.1). Note that some image phasecenters which were rejected in QL imaging are acceptable after being run through the SE imaging pipeline.

The QA process for coarse cubes has not yet been fully developed, but will be established when the production of coarse cubes begins in earnest in late 2026/early 2027. In general, it is expected that the QA process for individual polarization/spectral window planes will look very similar to the QA process for SECIs.

### 4.1 Image artifacts

Causes of image artifacts that require action are: "primary beam holes", high weights in the primary beam, RFI, and "bad baselines". The latter three types of artifacts require manual flagging, which typically happened at the QL imaging QA stage. The updated flagging is saved in the archive such that any calibration restore of the data contains these flags. Therefore, less flagging is required at the SE imaging stage given that flags from the QL stage have already been applied. The four main types of artifacts discussed in this section are shown in Figure 5.

Primary beam holes (PBHs) are due to the interaction of the OTF observing mode with CASA's `statwt` procedure. `Statwt` is run in the VLA calibration pipeline (in task `hifv_statwt`), and looks for sharp changes in amplitude with time or frequency. Such data spikes in time or frequency are down-weighted, as this is typically a symptom of RFI. However, in OTF mode the telescope rasters over the sky; every time it scans over a very bright source there is naturally a sharp change in amplitude with time that as a result is down-weighted for OTF data specifically. Because in OTF mode a particular time is associated with a particular location on the sky, this down-weighting leads to a low value in the mosaic primary beam at the affected location. The imaging pipeline masks values in the primary beam that are $< 0.2$ (where the maximum of the primary beam sensitivity is normalized to unity) so such down-weighted regions appear as "holes" (see top left panel of Fig. 5) and are not used in the deconvolution. As a result, the bright source causing the hole cannot be cleaned. For QL imaging, subimages (the final $\sim 1$ square degree image) with such PBHs were always rejected. However, for SE imaging, PBHs are not an issue: the SECI pipeline initially re-weights the data so that the weights are essentially uniform before running self-calibration; weights are then recalculated using residual data after a model image is made. As a result, PBHs due to bright sources do not appear in SE images. In QL images, however, two automated methods were developed to identify PBHs. One method is algorithmic and the other is a convolutional neural network (CNN). The algorithmic method identifies masked areas with "spikes" coming off them (deconvolution artifacts from the bright source that is causing the PBH) in the subimage which raise the overall noise of the image at that location. The CNN utilizes the TensorFlow Python package with a training set of 700 simulated PBH images and a validation set of 300 images. The CNN is 5 layers deep with kernels of $2 \times 2$ and a stride of 1; it identifies non-PBH images with a success rate of 99% and identifies true PBHs with a success rate of 79%. Missed PBH images are edge cases where the PBH barely protrudes into the $1 \times 1$ square degree subimage. Although the CNN does not seem to outperform the algorithmic method, both methods were retained for completeness.

A second type of artifact, which appeared in just a handful of QL images, is known as "high weights". Somewhat the opposite of primary beam holes, high weights are falsely high-weighted regions in the primary beam mosaic. It is unknown why this occurs. Affected images are identified quite easily algorithmically as the normalization of primary beam image leads to a value of unity for the highly weighted region, and thus significantly lower values for the rest of the primary beam (see top right panel of Fig. 5). Otto's high weights detection algorithm identifies primary beam images whose median sensitivity is $< 0.8$. Such images are set aside for manual flagging, and then re-imaged with the updated flagging such that the high weights in the primary beam no longer appear.

RFI became a more common problem over the lifetime of the survey. Like other VLA data that is processed by pipeline, VLASS depends on the standard pipeline RFI flagging using the CASA `rflag` procedure (an option in the `flagdata` task) that flags in both time and frequency space. Such flagging happens in both the calibration pipeline and the imaging pipeline. Even so, some RFI makes it through to the images. Sporadic, short-timescale RFI is fairly straightforward to identify in VLASS images because of the OTF observing mode: while rastering over the sky, a burst of RFI at a specific time is localized to a particular field or fields in the image, causing high noise over a region roughly the size of the telescope's field of view (see bottom left panel of Fig. 5). To identify such cases automatically, Otto looks for such high-noise regions. The data can then be manually flagged at the affected time and frequency.

The final type of common artifact that requires flagging is referred to colloquially as "bad baselines". The symptom is a series of stripes across the image (see bottom right panel of Fig. 5), stemming from a peak in the uv-data, which can be identified as a high peak in the Fourier transform of the affected image. Otto's bad baseline detector has evolved over the course of the survey. Originally it identified peaks in the Fourier transform of each image. The later version identifies peaks in the Fourier transform of the residual image which appears to be more effective in that it leads to fewer false positives.

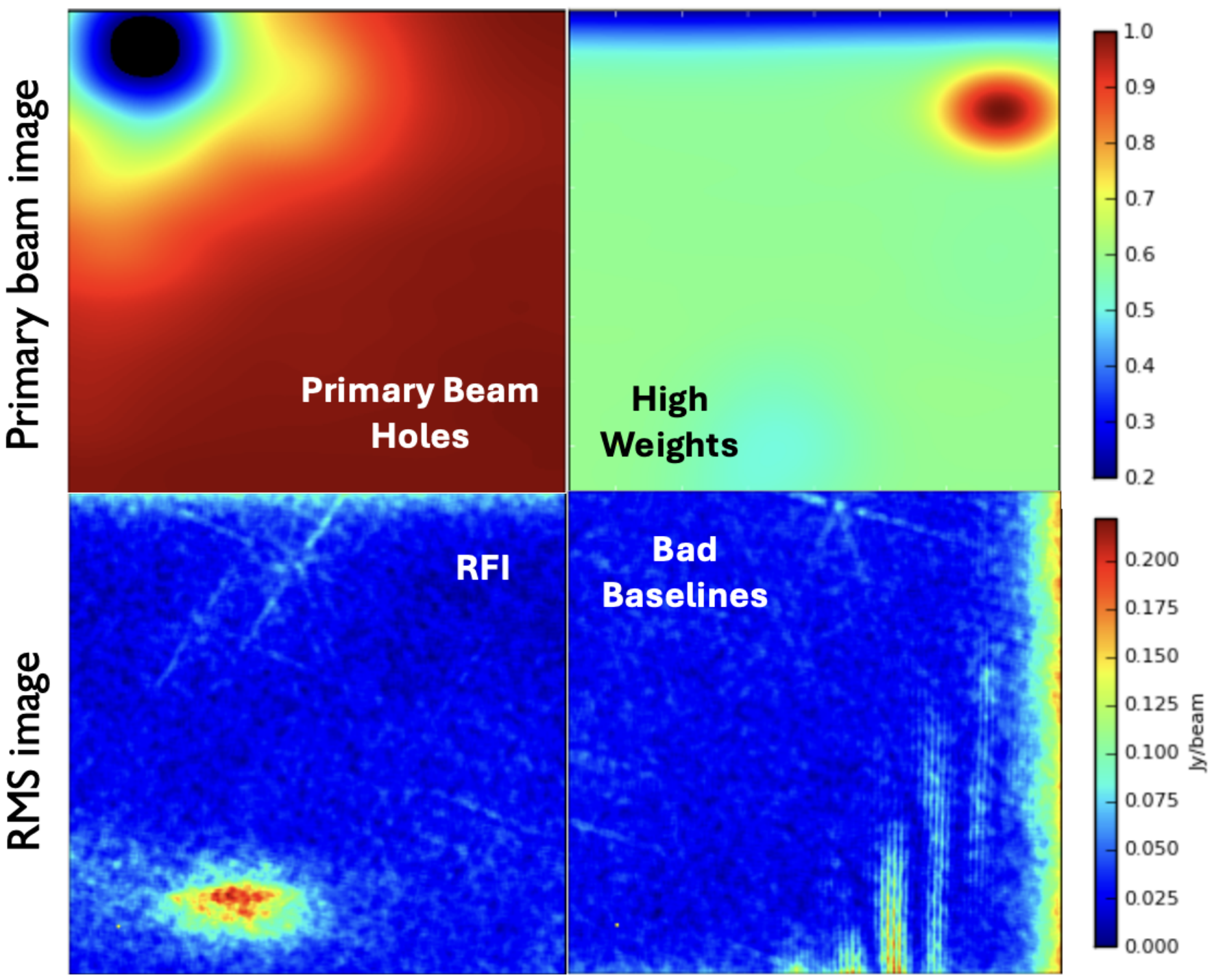

Figure 5. Image artifacts from VLASS that require flagging or image rejection. The top row shows examples of "primary beam holes" (PBHs) and high weights in primary beam images. The bottom row shows examples of RFI and "bad baselines" in RMS images. The examples shown are from Quick Look (QL) image products.

We note that additional artifacts may appear in the images due to poor deconvolution, such as spikes around bright sources, which do not respond to flagging. These artifacts can raise the overall RMS of an image, causing it to fail the statistical criteria for acceptance. Often reprocessing will produce an acceptable image. However, extremely bright sources can produce such strong deconvolution errors that no acceptable version was produced; such images are QA rejected. Over the entire survey, $< 0.6\%$ of images have been rejected.

## 5. DATA PROCESSING

Data processing for VLASS uses two 30-node clusters located in Socorro, New Mexico. Calibration took a relatively small amount of compute (around 200,000 core hours), but the imaging tasks are much more CPU intensive. QL imaging for 3.5 passes of the sky took $\approx 6$ million core hours. SECI and cube imaging will take much more processing (Table 2).

In response to the need for large amounts of imaging using $w$-projection, as discussed in Sec. 3.3, the CASA team developed two new gridders that include $w$-projection and also the antenna response ($a$-projection). The awp2 gridder uses CPUs and the awphpg gridder uses GPUs. The GPU-based gridder performs much faster, especially for $\gtrsim 8$ w-planes (where the speed of a 11-core CPU-based processing is comparable to that using a single NVIDIA L4 GPU). Although very low elevation fields will need 32 $w$-planes, fields at intermediate elevations can be processed with fewer planes and remain within the survey requirements and NRAO will probably used CPU-based processing for around half of the images needing w-projection as there is limited access to GPUs in house. We plan to begin processing the images needing $w$-projection after validation of the new gridders and incorporation into the VLASS workflow manager.

| Parameter | SE-continuum* (no w-projection) | SE-continuum (GPU) (w-projection) | SE-Continuum (CPU) (w-projection) | SE-Cube | Total |
|---|---|---|---|---|---|
| Number of jobs | 28,000 | 30,000 | 30,000 | 120,000 | 220,000 |
| Total CPU hrs | 2.5 million | 2.5 million | 28 million | 36 million | 79 million |
| Total GPU hrs | - | 5 million | - | - | 5 million |
| Total data in | 0.28 PB | 0.25 PB | 0.25 PB | 1PB | 2 PB |
| Total products | 28 TB | 25 TB | 25TB | 2 PB | 2.1 PB |
| RAM/CPU core | 64 GB | 128 GB | 16 GB | 30 GB | - |
| CPU Cores/job† | 1 | 1 | 11 | 4 | - |
| GPUs /job (L4) | - | 1 | - | - | - |
| Job duration‡ | 6 days | 3.5 days | 3.5 days | 3 days | - |
| Storage/job | 150 GB | 150 GB | 150GB | 170 GB | - |

Table 2. Remaining processing needs for VLASS, including a 13% contingency for reruns and assuming 50% of the w-projection imaging is via CPUs (awp2) and 50% via GPUs (awphpg) using 32 w-planes for awphpg and 8 w-planes for awp2. See Sec. 5.
* Excludes the $\approx 32,000$ images already made at the time of this publication.
† Core numbers can be varied for CPU-based jobs.
‡ Job durations can run 2-3 times longer on complicated fields with bright emission.

## 6. DATA ACCESS

Raw VLASS visibilities, like all science observations with the VLA, were immediately ingested into the NRAO Data Archive[††]. VLASS data have no proprietary period and thus were made available to the public within minutes of completion of each observation. Observations can be found in the NRAO Data Archive by searching for the name of the epoch or half epoch, e.g., "VLASS1.2" or "VLASS2". Calibration products are available for all observations which passed Quality Assessment (see Sec. 4). Observations with calibrations are identified by an icon and link under the "Cals" column of the search results. As was noted in Sec. 3.1, there are observations for which one version of calibration was used for QL images and another is being used for SE images. For any such observation, the calibration product labeled "1" in the archive is the most recent, SE version. Calibration products labeled "2" represent older versions of the calibration. We recommend always using the most recent, "1" calibration where multiple versions exist in the archive. As is typical for VLA observations, the raw data and the calibration products—rather than the calibrated data itself—are stored in the NRAO Data Archive. The raw data can be downloaded, the calibration products can be selected separately for download, and/or the restored calibration can be downloaded in which case the archive workflow uses CASA to apply the calibration products to the raw data before making the calibrated measurement set available to the user.

QL images and SECIs are available from NRAO in two locations. The first location is the NRAO Data Archive, where most published images can be found. However, as noted in Sec 3.2, the QL images for Epoch 1 that are in the archive have not been position-corrected for $w$-terms at the time of this publication. Position-corrected images can be found in the second location (see below). Note that where a SECI is available, NRAO always recommends using the SECI rather than the QL image. When SE coarse cubes begin production, they will be stored in the archive immediately upon being QA accepted. Note that images in the archive can be viewed through the archive using the Cube Analysis and Rendering Tool for Astronomy (CARTA).[13]

The second location where images can be found is through an NRAO server pointing directly to the images on disk.[‡‡]. The Quick Look directory contains the Epoch 1.1v2 and Epoch 1.2v2 QL images that have been corrected for $w$-terms. There is also a directory containing all SECIs, which is updated nightly with any new SECIs that were QA accepted that day. Due to disk space concerns, SE coarse cubes will not be kept long-term on disk at NRAO, but will be available in the archive as mentioned above.

The Canadian Astrophysics Data Center (CADC) also hosts the VLASS QL and SECI images in its VLASS collection and provides Virtual Observatory services to allow their download via scripted Application Program

[††]http://data.nrao.edu
[‡‡]https://archive-new.nrao.edu/vlass/

Interfaces. Finally, median stacked QL images from Epochs 1, 2 and 3 are available both from the NRAO server and as a Hierarchical Progressive Survey (HiPS)[14] in Aladin[15] or CARTA.

## 7. SUMMARY

The VLA Sky Survey (VLASS) is the highest resolution ($\sim 2.5$ arcsec) full sky ($\delta > -40^o$) radio survey to date. A synoptic survey, VLASS covered the sky 3.5 times from 2017 to 2026 using the VLA's on-the-fly (OTF) observing mode in the frequency range 2–4 GHz. A previous manuscript[1] focused on the science case for VLASS, while this manuscript focused on the data products, including generation, quality assurance (QA), and data access. Completed data products include calibration of the uv-data and "Quick Look" images for the entire survey. Other data products are in process: higher fidelity "Single Epoch" image products include continuum images (SECIs) as well as coarse cubes in polarization (IQU) for each 128-MHz spectral window. When SE processing of each epoch is complete, cumulative images will be made available. Catalogs are being produced as well. The creation of SECIs poses some difficulties, most particularly the requirement to deal with $w$-terms due to sky curvature. Roughly half of the sky needs consideration of $w$-terms, which requires a more sophisticated and computationally intensive algorithm in CASA, in order to obtain spectral indices within survey spec. Images are accessible in the NRAO data archive as well as on disk at NRAO at a location specified in Section 6.

## ACKNOWLEDGMENTS

The National Radio Astronomy Observatory is a facility of the National Science Foundation operated under cooperative agreement by Associated Universities, Inc.

## DISCLOSURES

The authors declare that there are no financial interests, commercial affiliations, or other potential conflicts of interest that could have influenced the objectivity of this research or the writing of this paper.